\documentclass[aps,prb,twocolumn,groupedaddress,amsmath,amssymb]{revtex4}

\usepackage{amssymb}
\usepackage{amsmath}
\usepackage{graphicx}
\usepackage{textcomp}
\usepackage{color}
\usepackage{amsfonts}
\usepackage{bbold}
\usepackage{dsfont}
\usepackage{epsfig}

\newcommand{\pd}[2]{\frac{\partial #1}{\partial #2}}

\newcommand{\arctanh}[1]{\operatorname{arctan}}

\begin{document}

\title{Electron transport across electrically switchable magnetic molecules}

\author{Sujeet K. Shukla}
\altaffiliation[Present Address: ]{Department of Physics, Indian Institute of Technology Delhi, New Delhi 110 016, India}
\affiliation{School of Physics and CRANN, Trinity College, Dublin 2, Ireland}
\author{Stefano Sanvito}
\email{sanvitos@tcd.ie}
\affiliation{School of Physics and CRANN, Trinity College, Dublin 2, Ireland}

\date{\today}

\begin{abstract}
We investigate the electron transport properties of a model magnetic molecule formed by two magnetic centers whose exchange 
coupling can be altered with a longitudinal electric field. In general we find a negative differential conductance at low temperatures 
originating from the different scattering amplitudes of the singlet and triplet states. More interestingly, when the molecule is strongly 
coupled to the leads and the potential drop at the magnetic centers is only weakly dependent on the magnetic configuration, we 
find that there is a critical voltage $V_\mathrm{C}$ at which the current becomes independent of the temperature. This corresponds 
to a peak in the low temperature current noise. In such limit we demonstrate that the quadratic current fluctuations are proportional 
to the product between the conductance fluctuations and the temperature. 
\end{abstract}

\maketitle


An intriguing aspect of electronic transport is the interaction between the current electrons and the internal 
degrees of freedom of the conductor. Atomic positions and vibrations are certainly at research center-stage, 
electro-migration being the most obvious example of interplay between the current and the atoms motion. 
The situation becomes even more intriguing at the nanoscale, where quantized vibrations can be detected 
by measuring the electron current and its derivatives with respect to the applied bias. This is the 
principle of inelastic electron tunneling spectroscopy (IETS). Furthermore also the reverse effect is possible, 
namely one can control the atomic positions of a nano-object by exciting appropriately some vibrational modes. 
Current-induced chemical reactions \cite{cicrs} and nano-catalysis \cite{nanocat} on surfaces are among the 
most appealing potential applications of this field.  

Equally important is the interplay between the electron current and the magnetic texture of a magnetic device. 
Such an interplay underpins the giant magnetoresistance effect \cite{gmr} and its reverse, i.e. current induced 
magnetization dynamics \cite{spintorque}. Considerably less investigated are the same phenomena at the 
atomic scale. This is mainly due to the intrinsic difficulties of both manipulating and detecting a few spins. 
In addition, magnetic excitations occur at energies lower than those involved in molecular vibrations, so that 
the measuring temperatures are often rather low. Still there are notable examples, such as ultra low temperature 
IETS of magnetic atoms on surfaces \cite{IBM} and of two-probe devices incorporating single magnetic 
molecules \cite{WenMS}. 

A new exciting prospect for scaling down spin-dynamics to the atomic level may be given by the ability of 
manipulating the magnetic configuration of a molecule with an electric potential instead of an electric current. 
Electrically induced alteration of the exchange coupling has been already predicted for two-centers magnetic 
molecules \cite{ESCE} and nanowires \cite{Angw}, and it is essentially based on the fact that the Stark shift of
a magnetic object may depend on its magnetic state. This effect can be a crucial ingredient for the physical 
implementation of quantum computing based on spins \cite{SMMQC,AffronteNew}. 

An intriguing question is whether or not the dependance of the exchange coupling over an electrical potential in a 
magnetic molecule can be detected electrically. This is the goal of our letter where we investigate the 
current-voltage, $I$-$V$, curve of a two-terminal device incorporating a two-center magnetic molecule in which 
the exchange coupling changes with bias. Importantly we find that, in particular conditions of coupling between
the molecule and the electrodes, there is a critical voltage $V_\mathrm{C}$ at which the current becomes 
independent of the temperature. This is accompanied by a negative differential conductance (NDC) at low 
temperature originating from the difference in scattering amplitude of the different spin-states of the molecule.


\begin{figure}[ht]
\includegraphics[width=8.0cm, clip=true]{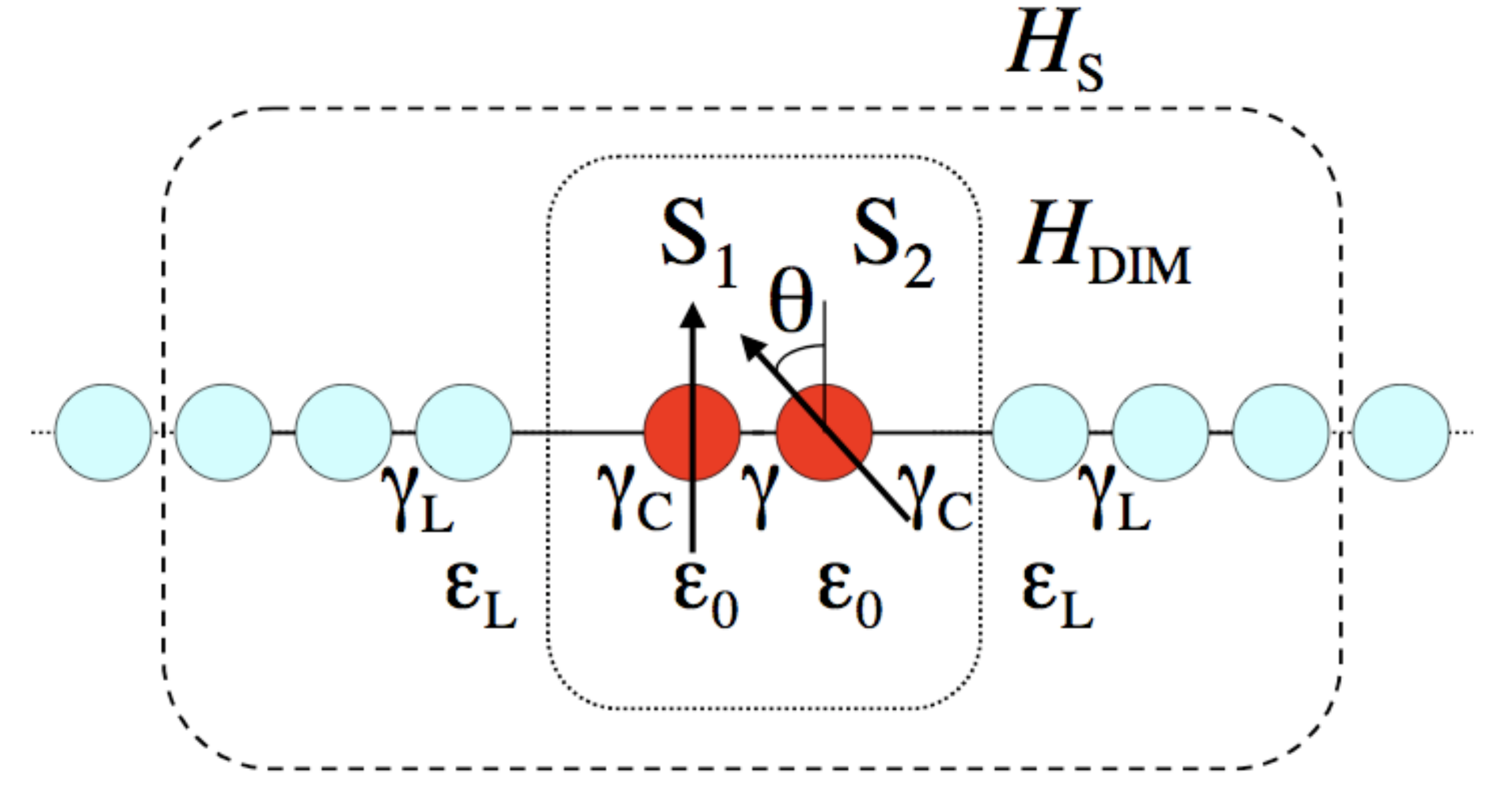}
\caption{(Color on line) The model system investigated: a dimer of magnetic atoms (red), carrying respectively
spin $S_1$ and $S_2$, is attached to two 1D non-magnetic electrodes (light blue). The Hamiltonian
for the dimer is $H_\mathrm{DIM}$. The scattering region (dashed box) includes the dimer and six atoms of the 
electrodes and it is described by the Hamiltonian matrix $H_\mathrm{S}$.}
\label{Fig1}
\end{figure}
In figure~\ref{Fig1} we show the simple model system investigated, which comprises a di-atomic magnetic
molecule sandwiched between two one-dimensional non-magnetic electrodes. The system is described
by the $s$-$d$ model \cite{Yosida96,Maria}, where spin of the current carrying $s$-electrons is exchange-coupled 
to the local spins $S_1$ and $S_2$ ($d$) of the two atoms in the dimer. $S_1$ and $S_2$ are treated 
as classical variables and their orientation determines the scattering potential for the $s$-electrons.
These are described by a tight-binding Hamiltonian with a single $s$-orbital per site at half-filling. The on-site energy 
and hopping integral in the electrodes are $\epsilon_\mathrm{L}=2$~eV and $\gamma_\mathrm{L}=-2$~eV, a choice 
which maintains the system far from the van Hoof singularities at any voltage investigated. No local spins are present 
in the electrodes so that their electronic structure is not spin-polarized. The Hamiltonian of the dimer is
\begin{equation}
H_\mathrm{DIM}=\sum_{\alpha\beta}\sum_i^{1,2}h_i^{\alpha\beta}c_{i}^{\alpha\dagger} c_{i}^{\beta}+
\sum_{\sigma}\gamma (c_1^{\sigma\dagger} c_2^{\sigma}+c_2^{\sigma\dagger} c_1^{\sigma})\:,
\end{equation}\label{Hdim}
where $h_i^{\alpha\beta}$ is the on-site Hamiltonian matrix of the $i$-th atom of the dimer, $\gamma$ is the hopping 
parameter and $c_i^{\sigma\dagger}$ ($c_i^{\sigma}$) is the creation (annihilation) operator for an electron
with spin $\sigma$ ($\uparrow,\downarrow$) at the site $i$. We have defined
$h_i^{\alpha\beta} = [\epsilon_0 + U(\rho_i-\rho_0)]\delta_{\alpha\beta} - J_\mathrm{sd}\vec{S}_i\cdot
\left({\vec{\sigma}}\right)_{\alpha\beta}$, where ${\vec{\sigma}}$ are the Pauli matrices, $\rho_i$ is the total occupation of 
$i$-th site, $\rho_0=1$ is the site occupation in the neutral configuration, $U=1$~eV is the atomic charging
energy and $J_\mathrm{sd}=2$~eV is the exchange parameter between the $s$-electrons and the local
spins. In our calculations we consider $\epsilon_0=2$~eV, $\gamma=-0.1$~eV, $|\vec{S}_i|=1$. 
In absence of spin-orbit interaction and spin-polarization of the electrodes the scattering potential is determined 
only by the mutual angle, $\theta$, between the two local spins. Finally, the dimer and the electrodes are coupled by 
the hopping integral $\gamma_\mathrm{C}$. In particular we explore the two cases in which 
$\gamma_\mathrm{C}=1/4\gamma_\mathrm{L}$ and $\gamma_\mathrm{C}=1/2\gamma_\mathrm{L}$. 
These parameters are only illustrative and have been chosen in order to maximize the difference in conductance 
between different spin-states of the molecule ($\theta=0$ vs $\theta=\pi$). 

The non-equilibrium Green function method \cite{Datta} applied to our tight-binding Hamiltonian \cite{Alex}
is used to calculate the transport properties. The central quantity is the retarded Green's function of the
scattering region 
\begin{equation}
G(E)= \lim_{\eta\to0+}\left[ (E+i\eta)-H_\mathrm{DIM}-\Sigma_\mathrm{L}-\Sigma_\mathrm{R}  \right]^{-1}\:,
\end{equation} 
where $E$ is the energy, $H_\mathrm{S}$ is the Hamiltonian matrix of the scattering region and 
$\Sigma_\mathrm{L}$ ($\Sigma_\mathrm{R}$) is the self-energy of the left- (right-) hand side electrode. 
This latter describes the interaction between the scattering region, which includes the dimer and 6 atoms of
the electrodes (see Fig.~\ref{Fig1}), and the electrodes. $G(E)$ enters in a self-consistent procedure to 
evaluate the stationary occupation of the scattering region and once convergence is achieved the 
two-probe {\it microscopic} current, $i(V)$, at the voltage $V$ is extracted from the Landauer formula \cite{Alex}.

Since at any given temperature the angle between the magnetic moments in the dimer fluctuates, for any
microscopic quantity $q$ we can define its {\it macroscopic} counterpart, $Q$, as the thermal average over all 
the possible angles
\begin{equation}
Q(V)=\langle q(V) \rangle=
\frac{\int_{{\cal E}_\mathrm{min}}^{{\cal E}_\mathrm{max}} q(\theta,V)\mathrm{e}^{-\frac{{\cal E}_{12}}{k_\mathrm{B}T}} \,\mathrm{d}{\cal E} }
{\int_{{\cal E}_\mathrm{min}}^{{\cal E}_\mathrm{max}} \mathrm{e}^{-\frac{{\cal E}_{12}}{k_\mathrm{B}T}} \, \mathrm{d}{\cal E}}\:,
\label{AvI}
\end{equation}
thus that if $q=i$ one obtains the macroscopic current, $I$.
Here ${\cal E}_{12}$ is the dimer magnetic energy, which writes
\begin{equation}
{\cal E}_{12}=-J_\mathrm{dd}\cos\theta\:,\;\;\;\;\;\;\;J_\mathrm{dd}=a+b\:v_\mathrm{d}^2(\theta,V)\:,
\label{DIMEN}
\end{equation}
and ${\cal E}_\mathrm{min}$ (${\cal E}_\mathrm{max}$) is its minimum (maximum) value.
In the equations (\ref{DIMEN}) above $J_\mathrm{dd}$ is the exchange energy between the 
two spins, which in turns is a quadratic function of the electrical potential difference between them, $v_\mathrm{d}$. This
latter is an intrinsic function of both $V$ and $\theta$. Finally the constants $a$ and $b$ are fixed to
the values of $a=0.001$~eV and $b=-0.8$~eV/V$^2$. Note that the functional dependance of $J_\mathrm{dd}$ 
over $v_\mathrm{d}$ implies a critical voltage at which the exchange energy changes sign, i.e. the magnetic 
coupling turns from ferromagnetic to antiferromagnetic\cite{ESCE,Angw}.

\begin{figure}[ht]
\includegraphics[width=9cm, clip=true]{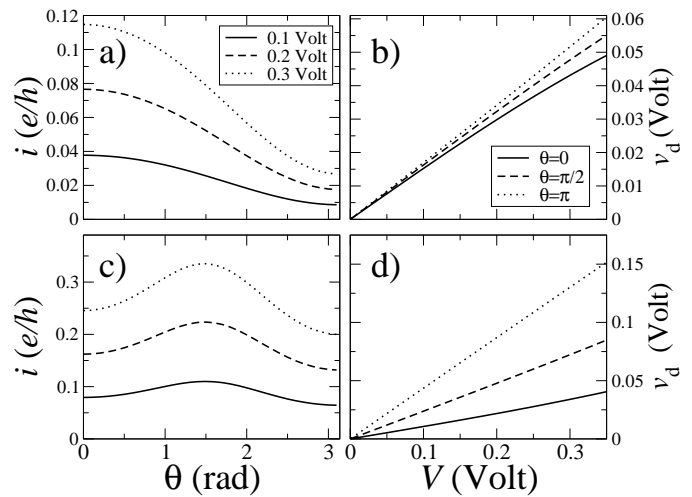}
\caption{Microscopic transport quantities. The microscopic current, $i$, as a function of the angle, $\theta$, is shown 
in panels (a) and (c) for different voltages $V$. Panel (b) and (d) show the internal potential drop, $v_\mathrm{d}$, 
as a function of the external bias and for different angles. Panels (a) and (b) are for 
$\gamma_\mathrm{C}=1/4\gamma_\mathrm{L}$, while (c) and (d) are for $\gamma_\mathrm{C}=1/2\gamma_\mathrm{L}$.}
\label{Fig2}
\end{figure}

We begin our analysis by investigating the microscopic quantities, i.e. the current $i(V,\theta)$ and the dimer internal 
potential drop $v_\mathrm{d}$. In Fig.~\ref{Fig2} we present, for both choices of coupling $\gamma_\mathrm{C}$, 
$i$-$\theta$ for different voltages and $v_\mathrm{d}$-$V$ for different angles $\theta$. In the case of 
$\gamma_\mathrm{C}=1/4\gamma_\mathrm{L}$ the current varies as $i(V,\theta)\sim [i_0+i_1\cos\theta]V$, with $i_0$ and $i_1$
two constants. At the same time $v_\mathrm{d}$ is only weakly dependent on the internal spin configuration, 
i.e. the $v_\mathrm{d}$-$V$ curve changes little with the angle $\theta$ [Fig.~\ref{Fig2}(b)].
In contrast for $\gamma_\mathrm{C}=1/2\gamma_\mathrm{L}$ the current peaks at approximately $\theta=\pi/2$ with 
both the parallel and antiparallel configurations being low conducting. Again the amplitude of the current variation over
$\theta$ increases with bias, although only moderately in this case. Furthermore for this situation $v_\mathrm{d}$,
which is still linear with $V$, is rather sensitive to the angle between the two spins. These differences affect dramatically 
the macroscopic current, $I$, that we calculate next. 

\begin{figure}[ht]
\includegraphics[width=9cm, clip=true]{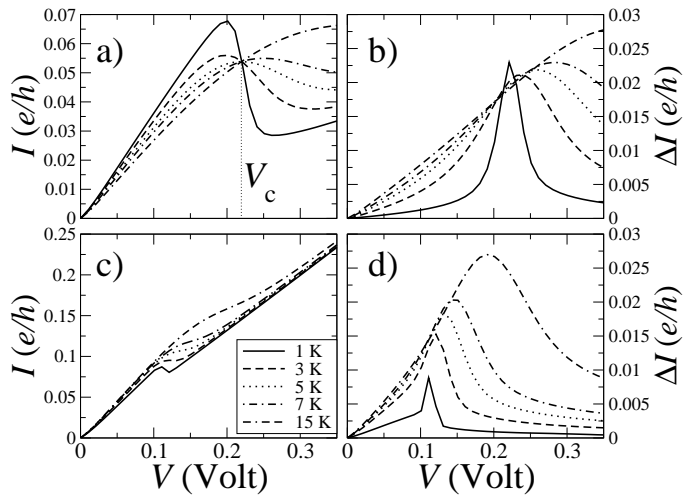}
\caption{Macroscopic transport quantities. The $I$-$V$ curves are presented in panels (a) and (c), while
panels (b) and (d) display the current quadratic fluctuations $\Delta I=\sqrt{\langle i^2\rangle-I^2}$ as a 
function of bias $V$. Panels (a) and (b) are for $\gamma_\mathrm{C}=1/4\gamma_\mathrm{L}$, while (c) and (d) are 
for $\gamma_\mathrm{C}=1/2\gamma_\mathrm{L}$. Note that in (a) there is a critical voltage $V_\mathrm{C}$ at which
the current becomes independent of the temperature.}\label{Fig3}
\end{figure}

The macroscopic $I$-$V$ curves for the two cases are presented in the panels (a) and (c) of Fig.~\ref{Fig3},
while the panels (b) and (d) report the current quadratic fluctuations $\Delta I=\sqrt{\langle i^2\rangle-I^2}$ 
still as a function of bias. In all cases we study the electrical response in the temperature range 1-15~K. 
The most interesting behaviour  is found for $\gamma_\mathrm{C}=1/4\gamma_\mathrm{L}$, from which 
we start our discussion. 


Figure~\ref{Fig3}(a) reveals two remarkable features. First we note that there is a pronounced NDC 
at about 0.2~Volt, which is well evident at 1~K, it weakens as the temperature increases and finally 
disappears at 15~K. Interestingly the scaling of the electrical current with the temperature is opposite at 
the two sides of the NDC: it decreases as the temperature is enhanced before the NDC while it grows with 
$T$ for voltages just after the NDC. The same NDC is present also in the case of stronger coupling
with the leads [$\gamma_\mathrm{C}=1/2\gamma_\mathrm{L}$, Fig.~\ref{Fig3}(c)] at the somewhat lower 
voltage of about 0.1~Volt. In this case however the NDC is much less pronounced and disappears already at 3~K. 

The second and most striking feature of Fig.~\ref{Fig3}(a) is the presence of a critical voltage, $V_\mathrm{C}$, at which
the current becomes independent of the temperature. Such a voltage is in the vicinity of the NDC and correlates
well with the peak in the current quadratic fluctuations [Fig.~\ref{Fig3}(b)] at low temperature. Note that this second feature 
is absent in the case of strong coupling to the leads. 

All these aspects can be easily understood by relating the microscopic quantities of Fig.~\ref{Fig2} with
the average of equation (\ref{AvI}). Let us consider the case of $\gamma_\mathrm{C}=1/4\gamma_\mathrm{L}$ first. 
In general the macroscopic current $I(V)$ is determined by the microscopic currents $i(V,\theta)$ of those 
configurations in which the system spends most of the time. The equations (\ref{DIMEN})
tell us that the ferromagnetic configuration is energetically favorable at low bias, while it is the 
antiferromagnetic to dominate at higher voltages (for $v_\mathrm{d}$ larger than $\pm\sqrt{-a/b}$). This
means that as the external bias increases the average current becomes progressively dominated
by antiferromagnetic configurations to the expenses of the ferromagnetic ones. Since the microscopic current
for $\theta=\pi$ is always considerably smaller than that for $\theta=0$ [see Fig.~\ref{Fig2}(a)],
this results in a decrease of the macroscopic current as a function of bias, i.e. in the NDC. Note that
this particular NDC is not of microscopic electronic origin since the microscopic currents
$i(V,\theta)$ are monotonic in $V$ for every $\theta$.

The fact that the exchange coupling changes sign as a function of the bias produces the second important 
feature in the macroscopic $I$-$V$ curve. In fact when the potential drop between the two magnetic atoms
is $v_\mathrm{d}=\pm\sqrt{-a/b}$, then the parallel and antiparallel configurations of the magnetic molecule
become energetically degenerate. This means that now no magnetic energy scale enters into the problem and
the system spends an equal amount of time in any spin configurations regardless of the temperature. 
In general $v_\mathrm{d}$ is proportional to the external bias $V$. Therfore one expects the existence of a 
universal external bias $V_\mathrm{C}=V[{\cal E}_{12}(v_\mathrm{d})=0]$ such that ${\cal E}_{12}=0$ and the macroscopic 
currents becomes independent of the temperature, as indeed demonstrated in Fig.~\ref{Fig3}(a). However there is a second 
condition for this to happen, i.e. $v_\mathrm{d}$ should be independent of the angle $\theta$. This is not 
satisfied for $\gamma_\mathrm{C}=1/2\gamma_\mathrm{L}$ [see Fig.~\ref{Fig2}(d)] 
and as a consequence the $I$-$V$ curves remain temperature dependent at any bias. 

%
\begin{figure}[ht]
\includegraphics[width=7cm, clip=true]{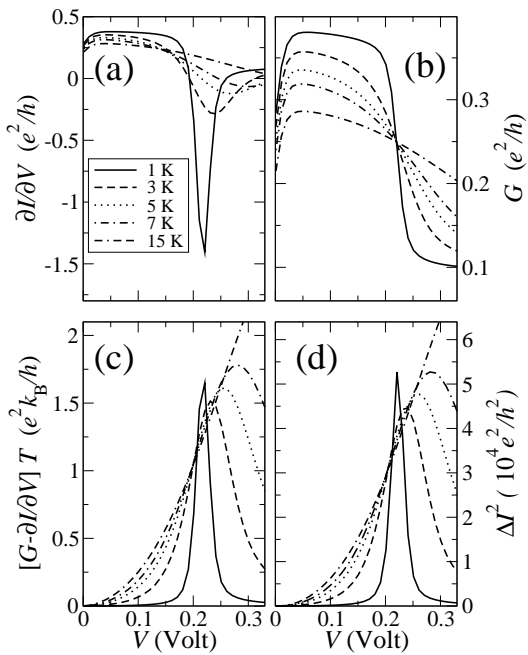}
\caption{Differential conductance calculated as (a) the derivative of the macroscopic current, $I$, with respect to the bias or
as the (b) thermal average, $G$, of the microscopic conductance. In panel (c) we show the product of $(G-\partial I/\partial V)$
with the temperature $T$, as a function of bias. Results are presented for $\gamma_\mathrm{C}=1/4\gamma_\mathrm{L}$. Note that 
$[G-\partial I/ \partial V]T$ is proportional to the square of the current quadratic fluctuations $\Delta I^2$ [panel (d)] of 
figure 3(b).}\label{Fig4}
\end{figure}
From our discussion it is now clear that if $v_\mathrm{d}$ is proportional to $V$ and weakly dependent on $\theta$, then
there will be a critical voltage $V_\mathrm{C}$ at which {\it any} macroscopic quantity becomes temperature independent. 
Figure~\ref{Fig3}(a) illustrates this feature for the current and the same is demonstrated in Fig.~\ref{Fig4}(b) for the conductance 
$G=\left \langle \pd{i}{V}\right \rangle$. Interestingly one can also adopt a different definition for the macroscopic conductance, namely 
that of the bias derivative of the macroscopic current $\partial I/\partial V$. Such a quantity is presented in Fig.~\ref{Fig4}(a) and as 
expected it appears sensibly different from $G$. Interestingly both $G$ and $\partial I/\partial V$ are, in principle, accessible from 
experiments, and one may wonder whether some general conclusions can be taken by measuring the two quantities independently. 

In general, by taking the equation~(\ref{AvI}) and formally deriving $Q$ with respect to the bias $V$ we find
\begin{equation}
\left[\left\langle\frac{\partial q}{\partial V}\right\rangle-\frac{\partial Q}{\partial V}\right]k_\mathrm{B}T=
\left\langle q\:\frac{\partial{\cal E}_{12}}{\partial V}\right\rangle-Q\left\langle \frac{\partial{\cal E}_{12}}{\partial V}\right\rangle\:,
\label{main1}
\end{equation}
where $k_\mathrm{B}$ is the Boltzman constant. If one now considers $q=i$ then the equation (\ref{main1}) establishes a general
relation between the conductance fluctuations and the correlation function between the current and the magnetic energy. Such a
relation is drastically simplified when the microscopic current has the typical spin-valve dependence $i(V,\theta)\sim [i_0+i_1\cos\theta]V$
and $v_\mathrm{d}$ is linear with $V$. In this situation (encountered here for $\gamma_\mathrm{C}=1/4\gamma_\mathrm{L}$) one finds
\begin{equation}
\left[G-\frac{\partial I}{\partial V}\right]k_\mathrm{B}T\propto\:\Delta I^2\:,
\label{main3}
\end{equation}
i.e. that the conductance fluctuations rescaled by the temperature are proportional to the squared current fluctuations. A numerical 
proof of such a relation is provided in the panels (c) and (d) of figure~\ref{Fig4}.


In conclusion we have investigated the temperature-dependent electronic transport through a model diatomic magnetic
molecule, in which the exchange coupling between the two magnetic centers is a function of the bias. This presents two 
remarkable characteristics. First, if the potential drop between the two magnetic centers is only weakly dependent on the angle 
between their magnetic moments and it is linear in $V$, then there is a critical voltage $V_\mathrm{C}$ at which the macroscopic 
current becomes temperature independent. Secondly, if in addition the microscopic current has a form $i(V,\theta)\sim [i_0+i_1\cos\theta]V$,
then there is a universal relation between the temperature-rescaled conductance fluctuations and the quadratic current fluctuations. 
Both these effects are a unique fingerprint of the dependance of the magnetic energy upon an external bias and can be
used as a tool for detecting such a dependence. 


This work is funded by Science Foundation of Ireland. We thank Maria Stamenova and Chaitanya Das Pemmaraju for helping with 
the numerical implementation and Tchavadar Todorov for useful discussion.

\end{document}